\documentclass[12pt]{article}
\usepackage{epsfig,amssymb,amsmath,psfrag}

\usepackage{pstricks}
\usepackage{color}

\def \be  {\begin{equation}}
\def \ee  {\end{equation}}
\def \ba  {\begin{eqnarray}}
\def \ea  {\end{eqnarray}}

\newcommand \Li{{\rm Li}}

\newcommand\bin[2]{\left({#1}\atop{#2}\right)}

\def \sha{{\,\amalg\hskip -3.6pt\amalg\,}}

\begin{document}

\thispagestyle{empty}
\null\vskip-12pt \hfill  LAPTH-033/12 \\
\null\vskip-12pt \hfill  CERN-PH-TH/2012-199 \\

\vskip2.2truecm
\begin{center}
\vskip 0.2truecm {\Large\bf
{\Large Generalised ladders and single-valued polylogs}
}\\
\vskip 1truecm
{\bf J.~M. Drummond  \\
}

\vskip 0.4truecm
{\it
CERN, Geneva 23, Switzerland\\
\vskip .2truecm                        }
and \\
\vskip .2truecm
{\it
LAPTH, Universit\'{e} de Savoie, CNRS\\
B.P. 110,  F-74941 Annecy-le-Vieux Cedex, France\\
\vskip .2truecm                        }
\end{center}

\vskip 1truecm 
\centerline{\bf Abstract} 

We introduce and solve an infinite class of loop integrals which generalises the well-known ladder series. The integrals are described in terms of single-valued polylogarithmic functions which satisfy certain differential equations. The combination of the differential equations and single-valued behaviour allow us to explicitly construct the polylogarithms recursively. For this class of integrals the symbol may be read off from the integrand in a particularly simple way. We give an explicit formula for the simplest generalisation of the ladder series. We also relate the generalised ladder integrals to a class of vacuum diagrams which includes both the wheels and the zigzags.

\medskip

 \noindent

\newpage
\setcounter{page}{1}\setcounter{footnote}{0}


\section{Introduction}

In perturbative quantum field theory, loop integrals play a central role. In order to study correlation functions or scattering amplitudes, the basic physical quantities of a quantum field theory, it is often the case that one needs efficient techniques for evaluating certain classes of diagrams in terms of explicit functions of the kinematical data, i.e. the particle momenta or positions. 

There exist powerful techniques for constructing loop integrands for scattering amplitudes (see e.g. \cite{Bern:1994zx,Bern:1994cg,ArkaniHamed:2010kv,Bourjaily:2011hi}). It is often the case that corresponding integrals and amplitudes evaluate to multi-dimensional polylogarithms, or iterated integrals.
Even at low loop orders, it is a non-trivial step to pass from this representation of the amplitude to one in which the multi-dimensional polylogarithms become explicit. 
An explicit representation is often desirable to allow straightforward analysis of physically interesting regimes, such as the OPE behaviour of correlation functions or Regge limits of scattering amplitudes, as well as for numerical evaluation.

When the relevant class of functions of the kinematical data is known in advance then it is often the case that quite powerful techniques can be applied to characterise an amplitude, or a given integral, in terms of functions from that class. An example that has been successfully employed in several recent works on amplitudes \cite{Goncharov:2010jf,Gaiotto:2011dt,Dixon:2011pw,Heslop:2011hv,Dixon:2011nj,Duhr:2012fh} is the notion of the symbol \cite{Chen,FBthesis,Gonch}, or more generally the Hopf structure \cite{Gonch2} associated to multi-dimensional polylogarithms. 

It would be very desirable, when given an expression in terms of loop integrals, to pass directly from the integrand to the symbol of the associated function. Here we will present a non-trivial class of integrals where the symbol can be read off immediately from the integrand in a very simple way.
One class of diagrams that has been known for a long time \cite{Usyukina:1993ch,Broadhurst:1993ib} is the set of ladder integrals. The integrals we will introduce are a natural generalisation of the ladder integrals and we will show how explicit solutions for the entire class can be obtained in terms of harmonic polylogarithms.

The use of differential equations in the study of loop integrals has a long history (for a discussion see e.g. \cite{Smirnov:2006ry}). The equations we study here are for off-shell integrals and are second-order inhomogeneous equations which relate $L$-loop integrals to $(L-1)$-loop integrals. They were used in \cite{Drummond:2006rz} to prove relations between different integrals. Similar equations for on-shell integrals were derived and studied in \cite{Drummond:2010cz}.

The solutions to the differential equations are not unique. An interesting feature of the class of generalised ladders is that the additional information required to select the correct solution comes from the fact that the associated functions must be single-valued. The single-valued polylogarithmic functions we obtain here are a family of generalisations of the Bloch-Wigner dilogarithm, which describes the simplest integral in the class of generalised ladders.

The map from the integrand to the symbol of the underlying function can be phrased quite succinctly in the language of the shuffle Hopf algebra. From the integrand one reads off a word constructed of letters drawn from the set $\{0,1\}$. Applying some basic Hopf algebra operations leads to a sum of products of harmonic polyogarithms. The function obtained from these operations agrees with the correct result up to terms containing explicit multi-zeta values multiplied by harmonic polylogarithms of lower degree. More precisely, the functions obtained obey the necessary differential equations but are not single-valued. In order to construct the extra terms one must look at the discontinuities of the obtained functions and ensure that the solutions of the differential equations are indeed single-valued.  

One interesting application of these results is in the study of vacuum integrals. In fact, the generalised ladder integrals we introduce here have finite two-point limits which are dual to the generalised zigzag vacuum diagrams studied in \cite{Dorynthesis}. This is an infinite family of vacuum diagrams which includes the wheel graphs and zigzag graphs. The zigzags are particularly interesting in that an all-loop formula was conjectured in \cite{Broadhurst:1995km} in terms of simple odd zetas only with a precise rational coefficient. They are also the only graphs in the class which appear as vacuum graphs of the $\phi^4$ theory. The methods discussed in this paper allow us to test this conjecture to quite high loop orders and will hopefully shed light on the simple structure of the transcendental numbers that appear.

The structure of this paper is as follows. In section \ref{sect-genlad} we review the structure of the ladder integrals and introduce the generalised ladders. We describe the differential equations satisfied by this class of integrals. In section \ref{sect-svp} we show that the relevant functions are single-valued polylogarithms and prove that this property fixes the solution of the differential equation uniquely. Then we introduce the concepts of the shuffle Hopf algebra and harmonic polylogarithms in section \ref{sect-shuffle} and use them to express a simple integrand-to-symbol map for the generalised ladders in section \ref{sect-solving}. We then discuss the problem of imposing the correct single-valued behaviour in section \ref{sect-svb}, showing that this can be achieved systematically in a recursive manner. In section \ref{sect-depth3} we give an explicit solution for the simplest examples of integrals in the class which are not ladders, the depth 3 integrals. In section \ref{sect-gensvpolys} we describe the relation of the problem to the Knizhnik-Zamolodchikov equation and the classification by Brown of single-valued polylogarithms. In section \ref{sect-4ptconf} we give a generalisation of the family of generalised ladders which is most natural in terms of four-point conformal integrals and for which the methods discussed in this article apply equally well. In section \ref{sect-vacuum} we discuss the relation to vacuum graphs and discuss a conjecture about the transcendental structure of the zigzag graphs due to Broadhurst and Kreimer.

\subsubsection*{Note}
This work was presented recently at Quantum Field Theory, Periods and Polylogarithms III, held in honour of David Broadhurst's 65th birthday. At this meeting I learned of closely related work by O. Schnetz \cite{os}

\section{Generalised ladder integrals}
\label{sect-genlad}

\begin{figure}
\centerline{{\epsfysize10cm \epsfbox{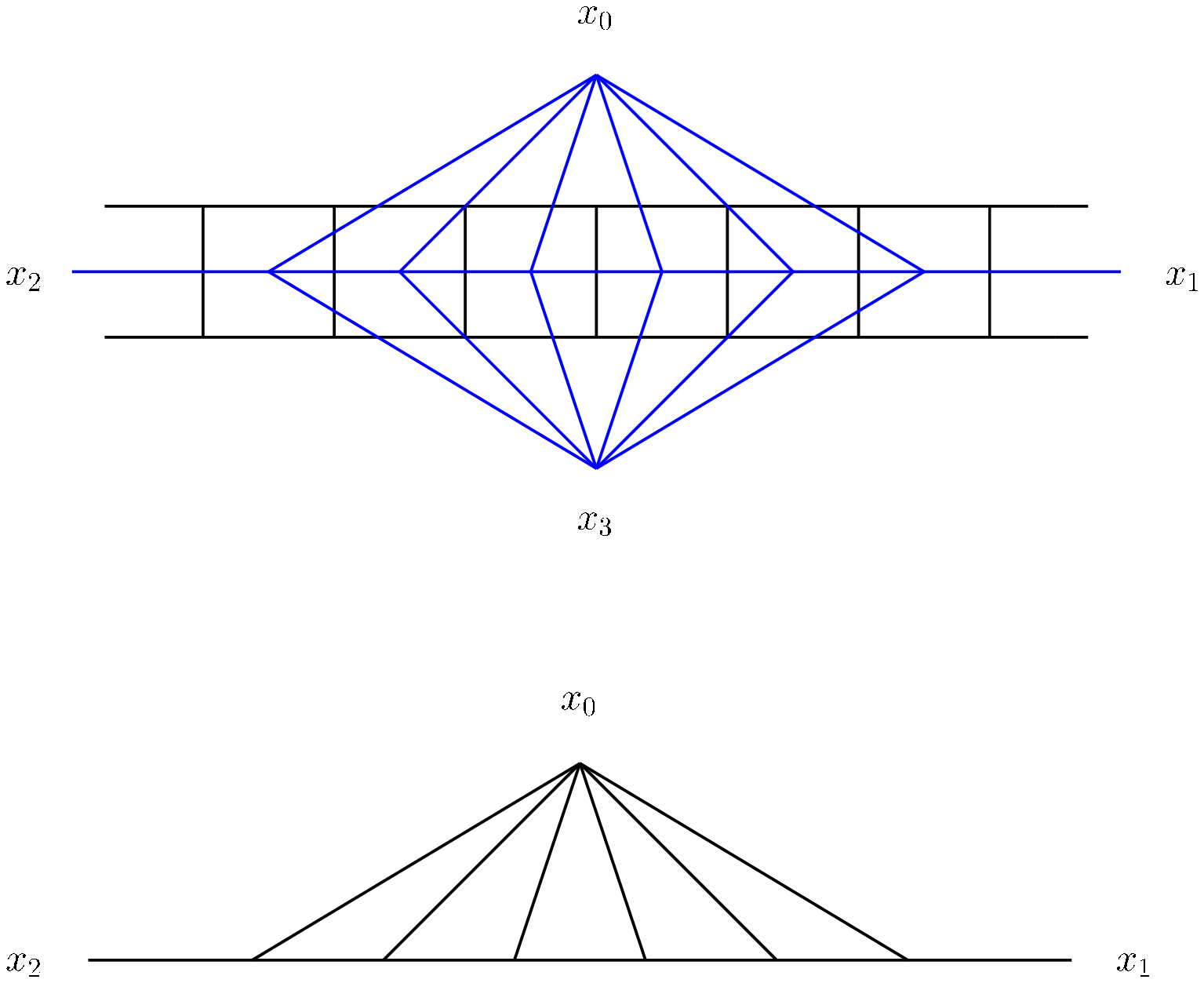}}} \caption[]{\small The four-point ladder integrals with the associated four-point and three-point dual diagrams.}
\label{ladders}
\end{figure}

Let us begin with the ladder integrals depicted in Fig.\ref{ladders}. We can represent the $L$-loop, four-point ladder integral in terms of dual coordinates in the following way,
\be
I^{(L)}(x_0,x_1,x_2,x_3) = \frac{1}{\pi^{2L}}\int \frac{1}{x_{b_1 2}^2} \prod_{i=1}^{L-1} \biggl(\frac{d^4x_{b_i} x_{03}^2}{\pi^2 x_{b_i b_{i+1}}^2 x_{b_i 0}^2 x_{b_i 3}^2}\biggr)\frac{1}{x_{b_L 0}^2 x_{b_L 1}^2  x_{b_L 3}^2}\,.
\label{4ptlad}
\ee
Here the integration vertices are labelled $x_{b_1}, \ldots , x_{b_L}$ while $x_{ij}^2$ denotes the square distance $(x_i - x_i)^2$ in four-dimensional Euclidean space. 
Since each integration vertex has four propagators attached the integral is conformally covariant and we may express the result as follows
\be
I^{(L)}(x_0,x_1,x_2,x_3) = \frac{1}{x_{03}^2 x_{12}^2} \Phi^{(L)}(s,t)\,.
\label{4ptladfn}
\ee
Here $s$ and $t$ are the two conformally invariant cross-ratios,
\be
s = \frac{x_{02}^2 x_{13}^2}{x_{12}^2 x_{03}^2}\,, \qquad t = \frac{x_{01}^2 x_{23}^2}{x_{12}^2 x_{03}^2}\,.
\ee
Note the numerators in (\ref{4ptlad}), which are not explicitly depicted in Fig.\ref{ladders},  ensure that the full integral has conformal weight 1 at each of the external vertices, in agreement with (\ref{4ptladfn}).
Since the integral is conformally covariant we lose no information by sending the point $x_3$ to infinity. Doing so we obtain the horizontal ladder diagram,
\be
I^{(L)}(x_0,x_1,x_2) = \frac{1}{\pi^{2L}}\int \frac{1}{x_{b_1 2}^2} \prod_{i=1}^{L-1} \biggl(\frac{d^4x_{b_i}}{\pi^2 x_{b_i b_{i+1}}^2 x_{b_i 0}^2}\biggr)\frac{1}{x_{b_L 0}^2 x_{b_L 1}^2}\,,
\ee
which is expressed in terms of the same function $\Phi^{(L)}$,
\be
I^{(L)}(x_0,x_1,x_2) = \frac{1}{x_{12}^2} \Phi^{(L)}(u,v)\,.
\ee
Here $u$ and $v$ are the limits of the cross-ratios as $x_3$ goes to infinity,
\be
u = \frac{x_{02}^2}{x_{12}^2}\,, \qquad v = \frac{x_{01}^2}{x_{12}^2}\,.
\ee

We will now describe a larger set of generalised ladder integrals.
Given a word $m = a_1 ... a_{L-1}$ with letters $a_i$ drawn from the set $\{0,1\}$, we will define a generalised $L$-loop, three-point ladder integral in the following way,
\be
I_m(x_0,x_1,x_2) = \frac{1}{\pi^{2L}}\int \frac{1}{x_{2 b_1}^2} \prod_{i=1}^{L-1} \biggl(\frac{d^4x_{b_i}}{x_{b_i b_{i+1}}^2 x_{b_i a_i}^2}\biggr) \frac{d^4x_{b_L}}{x_{b_L 0}^2 x_{b_L 1}^2}\,.
\ee
Here the integration vertices are labelled $x_{b_1}, \ldots , x_{b_L}$ while $x_{ij}^2$ denotes the square distance $(x_i - x_i)^2$ in four-dimensional Euclidean space. The case where all the $a_i$ are zero gives a three-point horizontal ladder integral\footnote{Strictly speaking these are the planar duals of the three-point horizontal ladder integrals.}. This series of integrals can be expressed in terms of classical polylogarithms \cite{Usyukina:1993ch}. The simplest example of all is where the word is empty and corresponds to the one-loop ladder integral (or triangle integral).

\begin{figure}
\psfrag{j}[cc][cc]{$j$} \psfrag{J}[cc][cc]{$J$}
\psfrag{j1}[cc][cc]{$j^{(1)}$} \psfrag{J1}[cc][cc]{$J^{(1)}$} 
\psfrag{pq}[cc][cc]{\!\!\!\!\!\!\!\!\!\!\!$p,q$}\psfrag{PQ}[cc][cc]{\,\,\,\,$P,Q$}
\psfrag{KS}[cc][cc]{\!\!\!\!\!\!\!\!\!\!\!$K,S$}\psfrag{ks}[cc][cc]{\,\,\,\,$k,s$}
\psfrag{rR}[cc][cc]{}
\psfrag{Tduality}[cc][cc]{T-duality} 
\centerline{{\epsfysize10.5cm \epsfbox{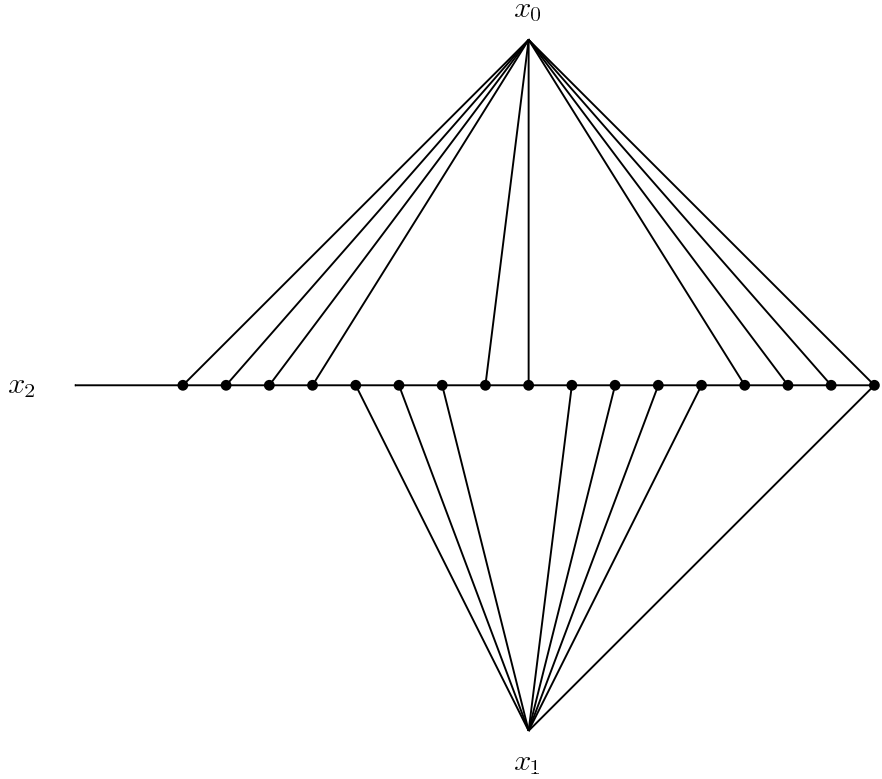}}} \caption[]{\small The generalised ladder integrals defined in dual coordinate space.}
\label{Fig:aux}
\end{figure}

Translation and rotation invariance together with the scaling dimension tell us that we can write the integrals as follows,
\be
I_m(x_0,x_1,x_2)  = \frac{1}{x_{12}^2} F_m(u,v)\,,
\label{function}
\ee
where $u$ and $v$ are given by
\be
u = \frac{x_{02}^2}{x_{12}^2}\,, \qquad v = \frac{x_{01}^2}{x_{12}^2}\,.
\ee

The generalised ladder integrals satisfy differential equations. This can be seen by applying the Laplace operator at the point $x_2$ and observing that we obtain a delta-function under the integral \cite{Drummond:2006rz},
\be
\Box_2 \frac{1}{x_{2b_1}^2} = - 4 \pi^2 \delta^4(x_2 - x_{b_1})\,.
\ee
We therefore find
\be
\Box_2 I_m(x_0,x_1,x_2) = -\frac{4}{x_{a_1 2}^2} I_{m'}(x_0,x_1,x_2),
\ee
where the word $m'$ is given by chopping off the first letter of $m$, i.e. $m' = a_2 ...a_{L-1}$. In the case where the word $m$ is empty the right hand side of the above equation becomes $-4/(x_{02}^2 x_{12}^2)$.
We can compare this with the action of the Laplace operator on the form (\ref{function}) and obtain \cite{Drummond:2006rz}
\be
uv\Delta^{(2)} F_m(u,v) = - c(a_1) F_{m'}(u,v)\,.
\ee
where $c(0)=1$ and $c(1)=u$ and the operator $\Delta^{(2)}$ is given by
\be
\Delta^{(2)} = 2(\partial_u + \partial_v) + u\partial_u^2 + v\partial_v^2 - (1-u-v)\partial_u \partial_v\,.
\ee
It is convenient to use the complex-conjugate variables $z$, $\bar z$ defined by
\be
u = \frac{z \bar z}{(1-z)(1-\bar z)}\,, \qquad v = \frac{1}{(1-z)(1-\bar z)}\,.
\label{defzzb}
\ee
In terms of these variables the differential equations become,
\be
d(a_1) \partial_z \partial_{\bar{z}} f_m(z,\bar{z}) = - f_{m'}(z,\bar z)\,.
\label{dezzb}
\ee
where $d(0) = z \bar z$ and $d(1) = (1-z)(1-\bar z)$. The functions $f_m(z,\bar z)$ are related to the functions $F_m(u,v)$ via
\be
f_m(z,\bar z) = \frac{z- \bar z}{(1-z)(1-\bar z)} F_m(u,v)\,.
\ee
Note that the $f_m$ are antisymmetric in the exchange of $z$ and $\bar z$ due to the antisymmetric prefactor and the symmetry of the variables $u$ and $v$. In the case that the word $m$ is empty the equation becomes.
\be
z \bar{z}(1-z)(1-\bar z) \partial_z \partial_{\bar z} f(z,\bar z) = -(z-\bar z)\,.
\label{1loopeq}
\ee

The differential equation (\ref{dezzb}) is not sufficient to fix $f_m$ in terms of $f_{m'}$. Indeed, taking into account the antisymmetry of $f_m$ we see that if $f^{\rm part}_m$ is a particular solution then the general solution is
\be
f_m(z,\bar z) = f^{\rm part}_m(z,\bar{z}) + h(z) - h(\bar z)
\label{ambiguity}
\ee
for an arbitrary holomorphic  function $h$.

\section{Single-valued Polylogs}
\label{sect-svp}

Let us consider the simplest case to see how the remaining freedom can be fixed. The one-loop ladder (or triangle) integral is given by 
\be
I(x_0,x_1,x_2) = \frac{1}{\pi^2} \int \frac{d^4x_b}{x_{0b}^2 x_{1b}^2 x_{2b}^2} = \frac{1}{x_{02}^2} \Phi^{(1)}(u,v)\,.
\ee
The function $\Phi^{(1)}$ can be expressed in terms of logarithms and dilogarithms \cite{'tHooft:1978xw,Usyukina:1992jd}. It is convenient to use the variable $z$ and $\bar z$ defined in (\ref{defzzb}), 
\be
\Phi^{(1)}(u,v) = \frac{(1-z)(1-\bar z)}{z-\bar z} \phi^{(1)}(z,\bar z)\,.
\ee
\be
\phi^{(1)}(z,\bar z) = 2(\Li_2(z) - \Li_2(\bar z)) + \log (z\bar z) (\log(1-z) - \log(1-\bar z))\,.
\label{Phi1t}
\ee
One may easily verify that the above function obeys the differential equation (\ref{1loopeq}),
\be
z \bar{z} (1-z)(1 - \bar z)\partial_z \partial_{\bar z} \phi^{(1)}(z,\bar z) = - (z - \bar z)\,.
\ee

In fact the function $\phi^{(1)}$ is (up to a factor of $4i$) the Bloch-Wigner dilogarithm function. This function has the special property that it is single-valued as we take $z$ around the points 0 or 1 in the complex plane. This is obvious from the above formula near $z=0$. It follows near $z=1$ from the symmetry $\phi^{(1)}(1-z,1-\bar z) = -\phi^{(1)}(z,\bar z)$. 
An alternative way to state the same property is that after an analytic continuation defined by treating the variables $z$ and $\bar z$ as independent complex variables in the formula (\ref{Phi1t}) the discontinuities around $z=0$ and $\bar z=0$  (and similarly around $z=1$ and $\bar z = 1$) are related,
\be
{\rm disc}_z \phi^{(1)} - {\rm disc}_{\bar z} \phi^{(1)} = 0\,, \qquad {\rm disc}_{1-z} \phi^{(1)} - {\rm disc}_{1- \bar z} \phi^{(1)} = 0\,.
\ee
This property is a special case of a more general phenomenon that the discontinuities of physical amplitudes are constrained by unitarity and locality of the underlying theory.

The criterion of single-valuedness selects a unique solution to the differential equation. Consider the general solution 
\be
f(z,\bar z) = \phi^{(1)}(z,\bar z) + h(z) - h(\bar z)\,.
\ee
Imposing that $f$ be single-valued implies 
\be
{\rm disc}_z h(z) = 0\,, \qquad {\rm disc}_{1-z} h(z) = 0\,.
\ee
From the integral formula it is clear that $h$ cannot have singularities at other values of $z$ which correspond to generic values of $\{ x_0,x_1,x_2 \}$. Moreover one can see that as any of the points $\{ x_0, x_1, x_2\}$ coincide or go to infinity one can see that the divergences are at worst logarithmic and so $h$ cannot have poles (including at $z=\infty$) and hence $h$ must be constant and therefore drops out in the combination $h(z)-h(\bar z)$.

More generally, the $L$-loop ladders are known \cite{Usyukina:1993ch,Isaev:2003tk},
\be
f_{0_{L-1}}(z,\bar z) = \phi^{(L)}(z,\bar z) = \sum_{r=0}^L  \frac{(-1)^r (2L-r)!}{L! r! (L-r)!} \log^r(z \bar z) (\Li_{2L-r}(z) - \Li_{2L -r}(\bar z))\,.
\label{Lloopladder}
\ee
and one can easily verify that they obey the differential equations (\ref{dezzb}) and are single-valued in the same sense as the Bloch-Wigner dilogarithm.

So generically we can see that the functions $f_m$ define a family of generalisations of the Bloch-Wigner dilogarithm, obeying the differential equations (\ref{dezzb}) and single-valued as we take $z$ around 0 or 1,
\be
{\rm disc}_z f_m - {\rm disc}_{\bar z} f_m = 0\,, \qquad {\rm disc}_{1-z} f_m - {\rm disc}_{1- \bar z} f_m = 0\,.
\label{svrels}
\ee

\section{The shuffle Hopf algebra and harmonic polylogs}
\label{sect-shuffle}

We will now introduce two concepts which are helpful to describe the generalised ladder functions. The first is a Hopf algebra based on shuffle relations. The second is a class of functions known as harmonic polylogs \cite{Remiddi:1999ew} which respect the shuffle product.

The shuffle algebra $\mathcal{A}$ is a commutative algebra generated by words of arbitrary length. The algebra has an identity element which is a generator of length zero. Given two words $w_1 = a_1 a_2 \ldots a_k$ and $w_2 = a_{k+1} a_{k+2} \ldots a_{l}$ the shuffle product $\sha$ is defined as the sum over all words of length $k$ preserving the orderings of the subsets of letters $a_1 , ... , a_k$ and $a_{k+1},\ldots,a_{l}$. That is we have
\be
a_1 \ldots a_k \sha a_{k+1} \ldots a_l = \sum_{\sigma} a_{\sigma(1)} a_{\sigma(2)} \ldots a_{\sigma(l)}
\ee
where $\sigma$ is a permutation obeying $\sigma^{-1}(a_i) < \sigma^{-1}(a_j)$ if $1\leq i <j \leq k$ or if $k+1 \leq i <j \leq l$. 
The algebra is naturally graded with the grading equal to the length.

The shuffle algebra admits a Hopf structure. That is we have a coproduct $\Delta : \mathcal{A} \longrightarrow \mathcal{A} \otimes \mathcal{A}$ which is given by the sum over all deconcatenations of a given word into two pieces,
\be
\Delta (a_1 \ldots a_l) = \sum_{j=0}^l a_1\ldots a_j \otimes a_{j+1} \ldots a_l\,.
\ee
The coproduct is coassociative
\be
({\rm id} \otimes \Delta) \Delta = (\Delta \otimes {\rm id})\Delta\,.
\ee

We also have the antipode $S : \mathcal{A} \longrightarrow \mathcal{A}$ which reverses a word and flips its sign for words of odd length
\be
S(a_1\ldots a_l) = (-1)^l a_l \ldots a_1\,.
\ee
We will sometimes use the notation $\tilde{w}$ for the reversed version of $w$.

We will now briefly describe a special class of iterated integrals, the harmonic polylogarithms \cite{Remiddi:1999ew}. Here we need only those integrals defined in terms of words with letters chosen from the set $\{0,1\}$. If the word is a string of $m$ zeros we have
\be
H(0_m;x) = \frac{1}{m!} (\log x)^m\,.
\ee
Otherwise we define
\begin{align}
H(0,w;x) &= \int_0^x \frac{dt}{t} H(w;t)\,,\notag \\
H(1,w;x) &= \int_0^x \frac{dt}{1-t} H(w;t) \,.
\end{align}
We take $H(1;x) = \int_0^x \frac{dt}{1-t} = - \log(1-x)$. Following common convention we absorb a string of $(p-1)$ zeros preceding a one by replacing the letter 1 by $p$, i.e. $H(0,0,1;x) = H(3;x)$. Sometimes we use the notation $H_w(x)$ for compactness.
The classical polylogs are a special case of the harmonic polylogs,
\be
{\rm Li}_n(x) = H(n;x)\,.
\ee

The harmonic polylogs satisfy the shuffle relation
\be
H(w_1;x) H(w_2;x) = H(w_1 \sha w_2;x).
\label{hplshuffle}
\ee
We will use the notation $\zeta(w) \equiv H(w;1)$ for $w$ a word made of zeros and ones. In the case that $H(w;1)$ is divergent we will use the shuffle algebra to separate off the divergent terms $H(1;1) = \zeta(1)$ and then define $\zeta(1)=0$. Multi-zeta values defined this way are often called shuffle zeta values and are sometimes denoted by $\zeta_\sha(w)$.
For words ending in a one this definition coincides with the usual definition of the multi-zeta values. For words ending in a zero, one may use the shuffle relations to relate $H(w;1)$ to the usual $\zeta$ functions with words ending in ones only. For example we have
\be
\zeta(2,0) \equiv H(2,0;1) = H(0;1)H(2;1) - 2 H(3;1) = -2\zeta(3)\,.
\ee

The fact that we have the shuffle relation (\ref{hplshuffle}) means that the harmonic polylogarithms provide a representation of the shuffle algebra on words built from zeros and ones. We can introduce the maps $\pi_x$ and $\tilde{\pi}_x$ which take a word $w$ to the harmonic polylogarithms corresponding to $w$ and $\tilde{w}$ respectively,
\be
\pi_x (w) = H(w;x)\,, \qquad \tilde{\pi}_x (w) = H(\tilde{w};x)\,.
\label{hplproj}
\ee

The harmonic polylogarithms are examples of more general iterated integrals which often appear in quantum field theory calculations. 
The functions come with a degree or weight, denoted $k$ below, and obey the property that
\be
d f^{(k)} = \sum_r f_r^{(k-1)} d \log \phi_r\,
\ee
where the sum over $r$ is finite and $\phi_r$ are rational functions and $f^{(0)}$ are rational constants. If $f^{(k)}(x)$ is a harmonic polylogarithm $H(w;x)$, the $\phi_r$ are always  $x$ or $(1-x)$. 

Given the above definition we can define the symbol via
\be
{\rm Sym} (f^{(k)}) = \sum_r {\rm Sym} (f^{(k-1)}_r) \otimes \phi_r\,.
\label{symboldef}
\ee
For the harmonic polylogarithms $H(w;x)$ the symbol can be read off from the word $w$ very simply by reversing the order of the word and replacing every 0 by $x$ and every 1 by $(1-x)$ and finally multiplying by $(-1)^d$ where $d$ is the number of 1 elements in the word $w$.
Since the symbol is so closely related to the word defining a given harmonic polylogarithm we will use the functions even when we want to describe the symbol of the function. Two combinations of harmonic polylogarithms of a given weight with the same symbol will differ only by terms involving explicit multi-zeta values (which drop out of the symbol as defined above).

\section{Solving the differential equations}
\label{sect-solving}

First we will give a simple way to construct a function with the same symbol as the $f_m$ describing the generalised ladder integrals. Given a word $m$ of length $(L-1)$ we form the word  of length $2L$ given by $w = m 01 S(m)$. 

\begin{figure}
\centerline{{\epsfysize3cm \epsfbox{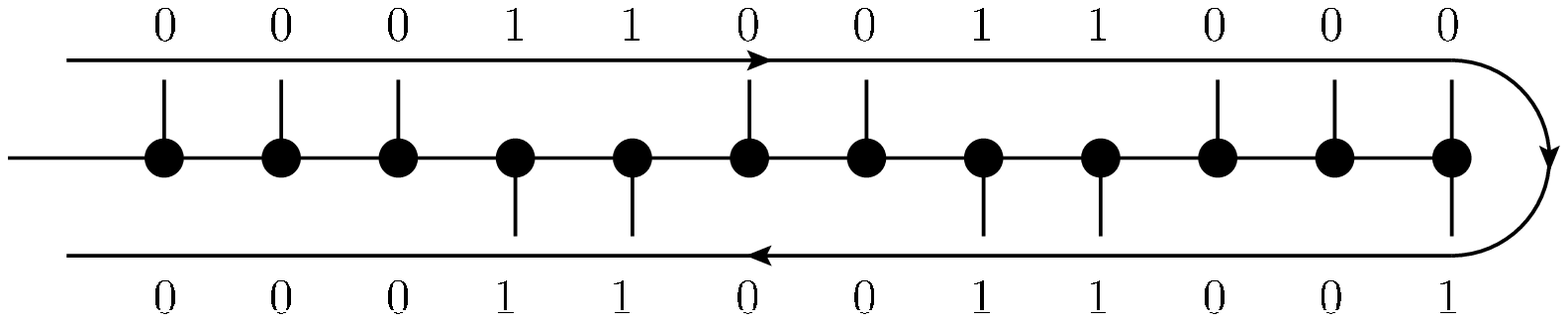}}} \caption[]{\small Reading off the word $w$ associated to a generalised ladder integral. To simplify the picture the propagators between the integration vertices and the external points have been shortened to stubs.}
\label{Fig:aux}
\end{figure}

Then we take all deconcatenations of $w = a_1 \ldots a_{2L}$,
\be
\Delta w = \sum_i a_1 \ldots a_i \otimes a_{i+1} \ldots a_{2L}\,,
\ee
and project on the left with $\pi_z$ and on the right with $\tilde{\pi}_{\bar z}$ as defined in (\ref{hplproj}). Then finally we subtract the same function with $z$ and $\bar z$ exchanged to obtain an antisymmetric function. Thus we define the `top part',
\be
f^{\rm top}_m(z,\bar z) = (\pi_z \otimes \tilde{\pi}_{\bar z}) \Delta w - (z \leftrightarrow \bar z)\,.
\label{ftopdef}
\ee
As we will see in section \ref{sect-gensvpolys},  the top part has an interpretation in terms of the Knizhnik-Zamolodchikov equation. We will refer to the number of ones in the word $w$ as the depth of the integral. The depth of the generalised ladders is odd as each one in the word $m$ counts twice and there is an additional 1 between $m$ and $S(m)$.

Let us see how the function $f^{\rm top}_m$ is constructed on some examples. First we consider the ladder case where $m$ is a string of $(L-1)$ zeros, $w= 0_{L-1}$. The antipode applied to $m$ gives the word $S(m) = (-1)^{L-1} 0_{L-1}$ thus the word $w$ of length $2L$ is
\be
w = (-1)^{L-1} 0_L 1 0_{L-1}\,.
\ee
Taking all deconcatenations and projecting as above gives us
\be
f^{\rm top}_{0_{L-1}}(z,\bar z) = (-1)^{L-1}\biggl[ \sum_{r=0}^{L-1} H_{L+1, 0_{L-1-r}}(z) H_{0_r}(\bar z) + \sum_{r=0}^L H_{0_r}(z) H_{L, 0_{L-r}}(\bar z) \biggr] - (z \leftrightarrow \bar{z})\,.
\label{altlad}
\ee
In fact the ladder case is particularly simple in that the function $f^{\rm top}_{0_{L-1}}$ is simply equal to the known ladder functions $\phi^{(L)}$(\ref{Lloopladder}). Note that the combinatorial coefficients in (\ref{Lloopladder}) have been absorbed into the way the zeros and ones are ordered in (\ref{altlad}).

Even without knowing the equivalence of $f^{\rm top}_{0_{L-1}}$ and $\phi^{(L)}$ for general $L$ we can see very simply that it is true for $L=1$. Moreover $f^{\rm top}_{0_{L-1}}$ obeys the differential equation,
\be
z \bar{z} \partial_z \partial_{\bar z} f^{\rm top}_{0_{L-1}}(z, \bar z) = - f^{\rm top}_{0_{L-2}}(z, \bar z)\,.
\ee
Then we can see that $f^{\rm top}_{0_{L-1}}$ obeys the single-valuedness criteria (\ref{svrels}) from the fact that
\be
{\rm disc}_x H_{a_1 \ldots a_k 1}(x) = 0, \qquad {\rm disc}_x H_{a_1 \ldots a_k 0_p}(x) = \sum_{r=1}^p H_{a_1 \ldots a_k,0_{p-r}}(x) \frac{(2 \pi i)^r}{r!}
\label{discx}
\ee
while 
\be
{\rm disc}_{1-x} H_{p}(x) = (2 \pi i) H_{0_{p-1}}(x)\,, \qquad  {\rm disc}_{1-x} H_{p,0_{k}}(x) = 0 \text{ for } k > 0\,.
\ee
This is another way to see that $f^{\rm top}_{0_{L-1}}$ must be the ladder function $\phi^{(L)} = f_{0_{L-1}}$.

Indeed in general we can see that $f^{\rm top}_m$ obeys 
\be
d(a_1) \partial_z \partial_{\bar z} f^{\rm top}_m(z,\bar z) = -f^{\rm top}_{m'}(z, \bar z)\,.
\ee
where $m=a_1\ldots a_{L-1} = a_1 m'$ and  $d(0)=z \bar z$ and $d(1) = (1-z)(1-\bar z)$. Note that the word $w$ of length $2L$ begins and ends with $a_1$, the first letter of $m$. This is why the prefactor of the differential operator $\partial_z \partial_{\bar z}$ only depends on $a_1$. The discontinuities can be checked using (\ref{discx}) and
\begin{align}
{\rm disc}_{1-x} H(a_1 \ldots a_k 1;x) &= (2\pi i) H(a_1 \ldots a_k;x) + {\rm MZVs}\,, \notag \\
{\rm disc}_{1-x} H(a_1 \ldots a_k 0;x) &= {\rm MZVs}\,,
\label{discomxmodMZV}
\end{align}
where `MZVs' means terms involving explicit multi-zeta values (i.e. terms with vanishing symbol in the sense of (\ref{symboldef}))\,. The equations (\ref{discx}) mean that $f^{\rm top}_m(z,\bar z)$ is single-valued around $z=0$, while (\ref{discomxmodMZV}) implies that it is single-valued around $z=1$ modulo terms involving explicit multi-zeta values. Thus we find that, modulo terms involving explicit multi-zetas, the function $f^{\rm top}_m$ is the correct solution to the differential equation for the generalised ladder functions. 
To obtain the full function $f_m(z,\bar{z})$ we will, in general, require additional terms containing explicit multi-zeta values to impose single-valued behaviour around $z=1$.

As an example we give the top part corresponding to the word containing a single 1, $m = 0_{L-p-2}10_{p}$. Following the procedure described above we find
\begin{align}
f^{\rm top}_{0_{L-p-2}10_{p}}(z,\bar z) = (-1)^{L-1}\biggl[\sum_{r=0}^{L-p-2} \!\! & H_{L-p-1,p+2,p+1,0_{L-p-2-r}}(z)H_{0_r}(\bar z) \notag \\
+ \sum_{r=0}^{p} &H_{L-p-1,p+2,0_{p-r}}(z)H_{L-p-1,0_r}(\bar z) \notag \\
+  \sum_{r=0}^{p+1} &H_{L-p-1,0_r}(z)H_{L-p-1,p+1,0_{p+1-r}}(\bar z) \notag \\
+ \!\!  \sum_{r=0}^{L-p-2} \!\! &  H_{0_r}(z) H_{L-p-1,p+1,p+2,0_{L-p-2-r}}(\bar z) \biggr] - (z \leftrightarrow \bar z)\,.
\label{ftopdepth3}
\end{align}
For this example we will go on to find the explicit multi-zeta terms needed to complete this function to the full $f_{0_{L-p-2}10_{p}}$ describing the associated generalised ladder integral. Before we do so we will explain some general conditions for imposing single-valued behaviour iteratively in the family of generalised ladders.

\section{Imposing single-valued behaviour}
\label{sect-svb}

We will now derive a set of constraints which will allow us to fix the ambiguity in solving the differential equation described in (\ref{ambiguity}). We will impose single-valued behaviour iteratively, that is we will assume that $f_{m'}$ is single-valued and then we will derive conditions which imply that $f_m$ is also single-valued. The two cases $m=0m'$ and $m=1m'$ are a little different so we proceed case by case.

\subsection{$a_1=0$}
Let us consider the differential equation (\ref{dezzb}) in the case that $a_1=0$ (i.e. $m=0m'$),
\be
z \bar{z} \partial_z \partial_{\bar{z}} f_m(z,\bar{z}) = -f_{m'}(z,\bar{z})\,.
\label{dea1=0}
\ee
We will suppose that all the $f_{m}$ are expressible in terms of a finite sum of products of harmonic polylogarithms evaluated at argument $z$ (an assumption that will be justified a posteriori). In this sum we will single out the terms which are proportional to a power of the logarithm of $z$ or $\bar z$, including the power 0, i.e. including any holomorphic or anti-holomorphic terms, so that for $f_{m'}$ we have
\be
f_{m'}(z,\bar{z}) = f_{m'}^{0}(z,\bar z) + \sum_{\alpha \geq 0} \bigl[g_{m'}^{\alpha}(z) H_{0_\alpha}(\bar z) - (z \leftrightarrow \bar z)\bigr]\,.
\label{fm'decomp}
\ee
Note that if the first letter of $m'$ is a 0, the term $f_{m'}^{0}(z,\bar z)$ is always simply obtained by integrating the corresponding function $f_{m''}^0$ from one loop lower, 
\be
f_{m'}^{0}(z,\bar z) = - \int_0^z \frac{dt}{t} \int_0^{\bar z} \frac{d \bar t}{\bar t} f_{m''}^{0}(t, \bar t)\,.
\ee
The case where $m' = 1 m''$ is even simpler. The function $f_{m'}^{0}$ is obtained by integrating the full function $f_{m''}$ from one loop lower,
\be
f_{m'}^{0}(z,\bar z) = - \int_0^z \frac{dt}{1-t} \int_0^{\bar z} \frac{d \bar t}{1-\bar t} f_{m''}(t, \bar t)\,.
\label{m'=1m''}
\ee

We will suppose further that $f_{m'}$ and all lower loop functions are single-valued functions. In particular we have for the analytic continuation of $f_{m'}$,
\be
({\rm disc}_z - {\rm disc}_{\bar z}) f_{m'}(z, \bar{z}) = 0\,, \qquad ({\rm disc}_{1-z} - {\rm disc}_{1-\bar z}) f_{m'}(z, \bar{z}) = 0\,.
\label{discsfw'}
\ee
The first of these conditions implies
\begin{align}
&({\rm disc}_z - {\rm disc}_{\bar z})f_{m',0}(z,\bar z) \notag \\
&+ \biggl[\sum_{\alpha \geq 0} \Bigl[ {\rm disc}_z g_{m'}^{\alpha}(z) H_{0_\alpha}(\bar z) - g^\alpha_{m'}(\bar z) \sum_{r=1}^\alpha H_{0_{\alpha -r}}(z) \frac{(2 \pi i)^r}{r!} \Bigr] + (z \leftrightarrow \bar{z}) \biggr]=0\,.
\label{fw'discz}
\end{align}
Moreover one can see iteratively that the first term above vanishes on its own. This is particularly clear in the case $m'=1m''$ described in (\ref{m'=1m''}). The function $f_{m''}$ is single-valued by assumption and the relation
\be
{\rm disc}_z \int_0^z \frac{dt}{1-t} g(t) = \int_0^z \frac{dt}{1-t} {\rm disc}_t \, g(t)\,
\ee
implies
\be
({\rm disc}_z - {\rm disc}_{\bar z})f_{m'}^{0}(z,\bar z) = 0\,.
\label{disczfm'0}
\ee
In the case $m'=0m''$ we find that
\be
{\rm disc}_z \int_0^z \frac {dt}{t} g(t) = \int_0^z \frac{dt}{t} {\rm disc}_t\, g(t)\,
\label{disczintdt/t}
\ee
implies (\ref{disczfm'0}) if $f_{m'',0}(z,\bar z)$ satisfies the same relation which we assume inductively.
Hence we find,
\be
({\rm disc}_z - {\rm disc}_{\bar z})f_{m'}^{0}(z,\bar z) =0\,, \qquad {\rm disc}_z g_{m'}^{\alpha} (z) = \sum_{\beta > \alpha} g_{m'}^{\beta}(z) \frac{(2 \pi i)^{\beta-\alpha}}{(\beta- \alpha)!}\,.
\label{fw'discz}
\ee
From the constraint at $z=1$ we find
\begin{align}
({\rm disc}_{1-z} - {\rm disc}_{1-\bar z})f_{m'}^{0}(z,\bar z) +\biggl[\sum_{\alpha \geq 0} {\rm disc}_{1-z} g^\alpha_{m'}(z) H_{0_\alpha}(\bar z)  - (z \leftrightarrow \bar{z})\biggr]=0\,.
\label{fw'discomz}
\end{align} 

Now we can write the general solution to the differential equation (\ref{dea1=0}) as follows
\be
f_m(z,\bar z) = - \int_0^z\frac{dt}{t} \int_0^{\bar z} \frac{d\bar t}{\bar t} f_{m'}^{0}(t,\bar t) + \sum_{\alpha \geq 0}\biggl[ \int_0^z \frac{dt}{t} g^\alpha_{m'}(t) H_{0_{\alpha +1}}(\bar z)  - (z \leftrightarrow \bar{z})\biggr] + h(z) - h(\bar z)\,.
\label{gensoln0}
\ee
Note that the holomorphic function $h$ corresponds to $g_m^0$ in the decomposition of $f_m$ analogous to the one for $f_{m'}$ given in (\ref{fm'decomp}).
Using the constraints (\ref{fw'discz}) and the relation (\ref{disczintdt/t})
we find that single-valuedness of $f_m$ around $z=0$ follows if
\begin{align}
{\rm disc}_z h(z)  = - \int_0^z \frac{dt}{t} \sum_{\alpha \geq 0} g^\alpha_{m'}(t) \frac{(2 \pi i)^{\alpha +1}}{(\alpha +1)!}\,.
\label{disczh}
\end{align}

To analyse the constraint of single-valuedness around $z=1$ we use (\ref{fw'discomz}) and the relation
\be
{\rm disc}_{1-z} \int_0^z \frac{dt}{t} g(t) = \int_1^z \frac{dt}{t} {\rm disc}_{1-t} g(t)\,.
\label{discomzg0}
\ee
We find
\be
{\rm disc}_{1-z} h(z) = \int_1^z \frac{dt}{t} \int_0^1 \frac{d\bar t}{\bar t} {\rm disc}_{1-t} f_{m'}^{0}(t,\bar t) = {\rm disc}_{1-z} \int_0^z \frac{dt}{t} \int_0^1 \frac{d\bar t}{\bar t} f_{m'}^{0}(t,\bar t)\,.
\label{discomzh}
\ee
The RHS of the final equality above can be rewritten as ${\rm disc}_{1-z} f_{m}^{0}(z,1)$ since the holomorphic functions $g_m^{\alpha}$ drop out after setting $\bar z =1$ and taking the discontinuity. It follows that (\ref{discomzh}) is equivalent to imposing
\be
{\rm disc}_{1-z} f_m(z,1) = 0\,.
\label{discfmz1}
\ee

The constraints (\ref{disczh},\ref{discomzh}) on its discontinuities are sufficient to determine the function $h(z)$ in terms of the lower loop function $f_{m'}(z,\bar z)$. We can give an explicit construction of $h(z)$ which will have the correct discontinuities. Firstly we note that if $g_{m'}^{0}$ has an expression in terms of harmonic polylogarithms as follows,
\be
g_{m'}^0(z) = \sum c_i H(w_i;z)\,,
\ee
then the following function,
\be
h_0(z) = \sum_i c_i H(0w_i0;z)
\ee
has the correct discontinuities around $z=0$, as given in (\ref{disczh}). Now, if we have a map $\mathcal{F}$ which takes a combination of harmonic polylogarithms and removes the discontinuity around $z=0$ while preserving the discontinuity around $z=1$, we can construct the function $h(z)$ as follows,
\be
h(z) = h_0(z) +\mathcal{F}\bigl[ f_m^0(z,1) - h_0(z) \bigr]\,.
\label{hformula0int}
\ee
We give an explicit construction of the map $\mathcal{F}$ in Appendix \ref{app-Fmap}.

\subsection{$a_1=1$}
Now let us consider the differential equation (\ref{dezzb}) in the case that $a_1=1$, i.e.
\be
(1-z)(1- \bar{z}) \partial_z \partial_{\bar{z}} f_m(z,\bar{z}) = -f_{m'}(z,\bar{z})\,.
\label{dea1=1}
\ee
In that case we may write the general solution as follows,
\be
f_{m}(z,\bar z) = -\int_0^z \frac{dt}{1-t} \int_0^{\bar z} \frac{d\bar t}{\bar t} f_{m'}(t,\bar t) + h(z) - h(\bar z)\,.
\label{a1=1intformula}
\ee
Since we have
\be
{\rm disc}_z \int_0^z \frac{dt}{1-t} g(t) = \int_0^z \frac{dt}{1-t} {\rm disc}_t \, g(t)\,,
\ee
imposing single-valued behaviour for $f_m$ at $z=0$ amounts to
\be
{\rm disc}_z h(z) = 0\,.
\label{disczh2}
\ee

Taking a discontinuity at $z=1$ requires a little care. In fact we have
\be
{\rm disc}_{1-z} \int_0^z \frac{dt}{1-t} g(t) = \lim_{\epsilon \rightarrow 0} \biggl[(2 \pi i) g(1-\epsilon) + \int_1^z \frac{dt}{1-t-\epsilon} {\rm disc}_{1-t} g(t)  \biggr]\,.
\label{divdisc}
\ee
The regularisation $\epsilon$ is needed here in the case that $g(t)$ diverges logarithmically as $t \rightarrow 1$, which happens when the first letter of $m'$ is also a 1. Of course the combination of both terms on the RHS above is finite in the limit as $\epsilon$ goes to zero. If $m'$ begins with a 0 then $g(t)$ is finite as $t\rightarrow 1$ and ${\rm disc}_{1-t}g(t)$ vanishes as $t \rightarrow 1$. This case is sufficient for all our purposes in this paper so to simplify the presentation we assume here that $m'$ begins with a 0. Imposing single-valuedness at $z=1$ for $f_m$ implies
\be
{\rm disc}_{1-z} h(z) = -(2\pi i) \int_0^z \frac{dt}{1-t} f_{m'}(t,1) + \int_0^z \frac{dt}{1-t} \int_0^1 \frac{d \bar{t}}{1-\bar{t}} {\rm disc}_{1-t} f_{m'}(t,\bar t)\,.
\label{discomzh2}
\ee
The formulae (\ref{disczh2},\ref{discomzh2}) define $h(z)$ in terms of $f_{m'}(z,\bar z)$. 

Just as in the case where $a_1=0$ we may make use of the $\mathcal{F}$ map to write an explicit expression for $h(z)$. We find
\be
h(z) =  \mathcal{F}\biggl[ \int_0^z \frac{dt}{1-t} \int_0^1 \frac{d \bar{t}}{1 - \bar{t}} f_{m'}(t,\bar t) - H_1(z) \int_1^z \frac{dt}{1-t} f_{m'}(t,1) \biggr]\,.
\label{hformula1int}
\ee
Here the divergent $\bar{t}$ integral in the first term is regulated by the shuffle regularisation and we have made use of the fact that ${\rm disc}_{1-z} f_{m'}(z,1) = 0$ which is the constraint (\ref{discfmz1}) applied to $f_{m'}$.

Using the methods outlined in this section we can automate the construction of the single-valued polylogarithms associated to the generalised ladders up to high loop order. For example the `zigzag' series $f_{1,0,1,0,\ldots}$ can easily be obtained up to $L=12$ (i.e. transcendental weight 24). In the next section we will give formulas for the depth 3 series to arbitrary loop order.

\section{The depth 3 family}
\label{sect-depth3}

Now let us explicitly construct the single-valued polylogarithms which represent the simplest non-ladder class of integrals, i.e. the depth 3 integrals. This two-parameter family of  integrals is represented by the words $m=0_{L-p-2}10_{p}$.
\begin{figure}
\centerline{{\epsfysize9cm \epsfbox{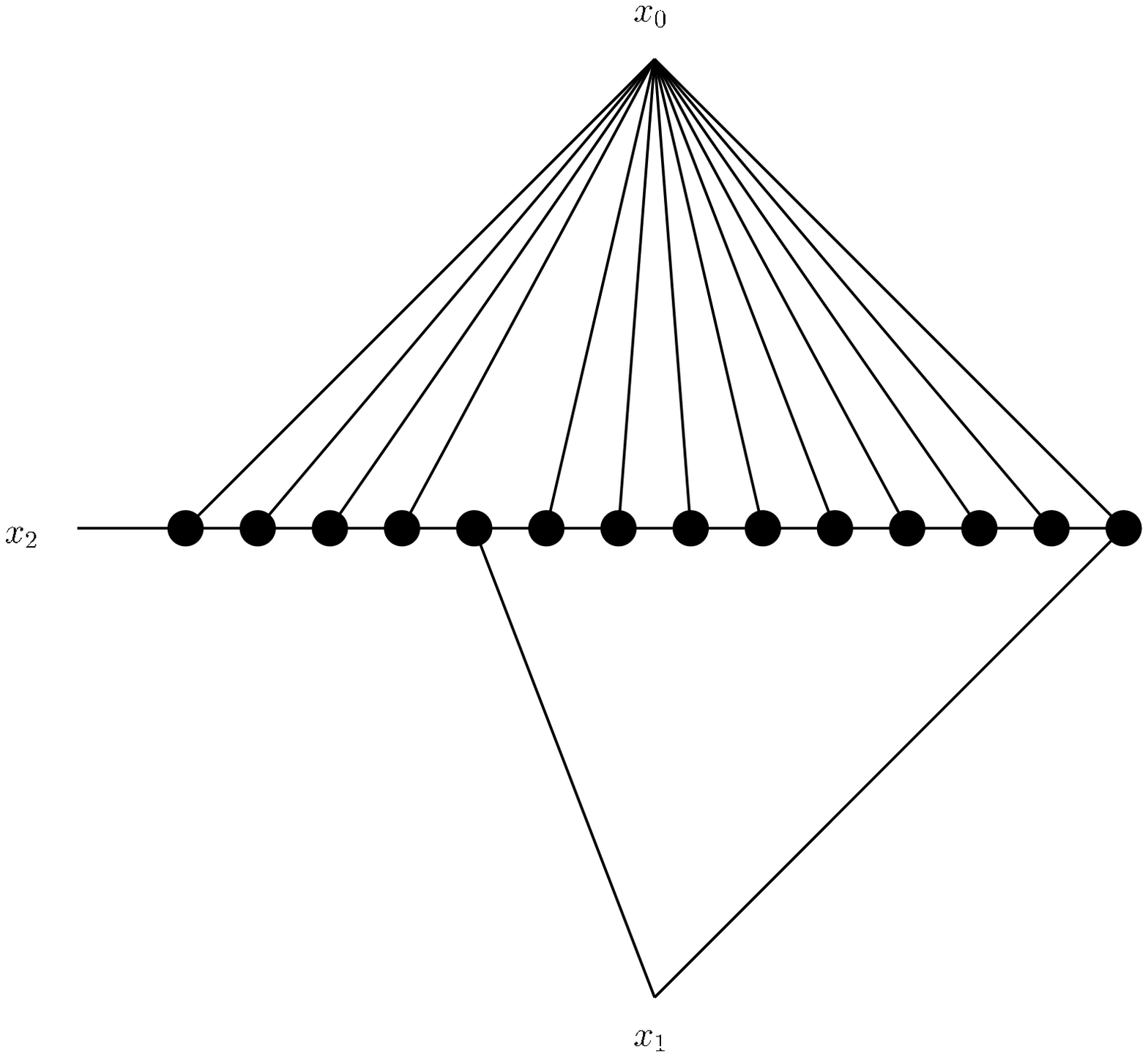}}} \caption[]{\small The depth 3 family of generalised ladders.}
\label{Fig:aux}
\end{figure}
We will write the full expression for this class of depth 3 generalised ladders as 
\be
f_{L-p-1,0_p}(z,\bar{z}) = f^{\rm top}_{L-p-1,0_p}(z,\bar{z})+f^{\rm ex}_{L-p-1,0_p}(z,\bar{z}).
\label{topexd3}
\ee
The function $f^{\rm top}_{L-p-1,0_p}$ is given explicitly (\ref{ftopdepth3}). Here we will construct the piece $f^{\rm ex}_{L-p-1,0_p}(z,\bar z)$.

We will start by considering the case $L=p+2$ so that the associated word reads $m=10_p$. In this case the associated differential equation reads
\be
(1-z)(1-\bar z) \partial_z \partial_{\bar z} f_{10_p}(z,\bar z) = - f_{0_p}(z,\bar z) = - \phi^{(p+1)}(z,\bar z)\,.
\ee
we see that the RHS is given in terms of the ladder function $\phi^{(p+1)}$ which is known explicitly. 
Now, on the one hand we have the integral formula (\ref{a1=1intformula}) and on the other the decomposition formula (\ref{topexd3}) separating the top part from the term containing explicit multi-zetas. 
Let us split the function $h(z)$ from the integral formula into a piece coming from $f^{\rm top}$ and a piece coming from $f^{\rm ex}$,
\be
h(z) = h^{\rm top}(z) + h^{\rm ex}(z).
\ee
Here we have 
\be
h^{\rm top}(z) = (-1)^p \bigl[H_{1,p+2,p+1}(z) - H_{1,p+1,p+2}(z)\bigr]
\ee
Comparing the two formulas (\ref{ftopdepth3},\ref{a1=1intformula}) we see that $f^{\rm ex}$ receives contributions only from the single-variable function $h(z)$.
\be
f^{\rm ex}(z, \bar z) = h^{\rm ex}(z) - h^{\rm ex}(\bar z)\,.
\ee
Now we need to impose the discontinuity constraints (\ref{disczh2},\ref{discomzh2}) on $h(z)$. Following the procedure outlined in section \ref{sect-svb} we find
\begin{align}
h^{\rm ex}(z) = &\sum_{m=0}^{\lfloor \frac{p-1}{2} \rfloor} 2 \bin{2(p-m)+1}{p+1} \zeta(2p+1-2m) H_{1,2m+2}(z) \notag \\
&- \bin{2(p+2)}{p+2} \zeta(2p+3)H_1(z) - 2 \delta_{p,{\rm odd}} H_{1,p+1}(z) \zeta(p+2)\,.
\label{firstd3h}
\end{align}
The final form for the single-valued polylogarithms associated with this class of of generalised ladder integrals is given by
\be
f_{10_p}(z,\bar z) = f_{10_p}^{\rm top}(z,\bar z) + h^{\rm ex}(z) - h^{\rm ex}(\bar z)\,.
\label{firstd3int}
\ee

We can continue to consider the most general depth 3 integrals. This requires solving the differential equation
\be
z \bar{z} \partial_z \partial_{\bar z} f_{L-p-1,0_p}(z, \bar z) = - f_{L-p-2,0_p} (z,\bar z)\,.
\ee
The RHS is given in terms of functions in the same class at one lower loop, with the initial term given by the functions (\ref{firstd3int}) that we have just solved for above. The general solution to the equation is given by (\ref{gensoln0}). Imposing single-valued behaviour as described in section \ref{sect-svb} we find the following structure for the result, 
\be
f^{\rm ex}_{L-p-1,0_{p}}(z,\bar z) = (-1)^{L-1} \sum_{\alpha =0}^{L-p-2}\Bigl(q^\alpha_{p,L}(z) H_{0_\alpha}(\bar z) - q^\alpha_{p,L}(\bar z) H_{0_\alpha}(z)\Bigr)
\ee
with
\begin{align}
q^{\alpha}_{p,L}(z) = \sum_{k=0}^{L-p-2-\alpha} \biggl[ &H_{L-p-1,p+2,0_{L-p-2-\alpha -k}}(z) c_{p,k} 
+H_{L-p-1,p+1,0_{L-p-2-\alpha-k}}(z) d_{p,k} \notag \\
+&\sum_{m=1}^{\lfloor \frac{p}{2} \rfloor} H_{L-p-1,2m,0_{L-p-2-\alpha-k}}(z) e_{p,k,m} 
+ H_{L-p-1,0_{L-p-2-\alpha-k}}(z) t_{p,k} \biggr]\,.
\end{align}

The fact that the coefficients $c,d,e,t$ do not depend explicitly on $\alpha$ and $L$ follows from imposing single-valued behaviour around $z=0$. Imposing single-valuedness around $z=1$ allows us to solve for the coefficients $c,d,e,t$ given the boundary case where $L=p+2$ above. We find $c_{p,k}=0$ if $k=0$ or if $k+p$ is odd. For the other coefficients we find $e_{p,k,m}=0$ for $k>0$, while $t_{p,k}=0$ for $k$ odd. The non-zero values are given by
\begin{align}
c_{p,k} &= 2 (-1)^{k+1} \bin{k+p}{p} \zeta(k+p+1), \notag \\
e_{p,k,m} &=  (-1)^p 2 \bin{2p+3-2m}{p+1} \zeta(2p+3-2m)\,, \notag \\
t_{p,k} &= (-1)^{p+1} 2\bin{2p+3+k}{p+1} \zeta(k+2p+3)\,,
\label{cdetcoeffs}
\end{align}
while $d_{p,0}=-c_{p,1}$ and $d_{p,k} = -c_{p+1,k}$ for $k>0$. Note that all coefficients above are proportional to simple odd zeta values. 
The full single-valued polylogarithm for the depth 3 case is
\be
f_{L-p-1,0_p}(z,\bar z) = f^{\rm top}_{L-p-1,0_p}(z,\bar z) + f^{\rm ex}_{L-p-1,0_p}(z,\bar z)\,.
\label{depth3full}
\ee
Thus we have found explicit single-valued solutions for an infinite two-parameter family of generalised ladder integrals. The methods of section \ref{sect-svb} can be used to extend this result to higher depth.

\section{Single-valued polylogs and the KZ equation}
\label{sect-gensvpolys}

We saw in section \ref{sect-solving} that the top part (\ref{ftopdef}), or equivalently the symbol in the sense of equation (\ref{symboldef}), of all the generalised ladders can be very easily constructed just by reading off a word from the diagram. From this labelling word $w$ we construct a word of length $2L$, given by $w 0 1 S(w)$. From this word of length $2L$ we translate directly to the symbol of the function $f_m$ by taking the sum over all deconcatenations and projecting onto harmonic polylogarithms of $z$ on the left half and harmonic polylogarithms of $\bar z$ on the reversed right half. 

To get from the top part to the full function we saw that in the case of the ladders we needed to do nothing extra while in the next simplest case, the depth 3 integrals, we had to carefully make sure that the differential equation was solved by a function obeying the criterion of single-valuedness. In the depth 3 case this meant that we had extra terms with explicit appearances of zeta values.

Thus we can capture the essence of the problem by saying we need single-valued polylogarithm functions satisfying the tower of differential equations (\ref{dezzb}). A similar problem is addressed in \cite{fbsvpl}\footnote{I would like to thank Claude Duhr for bringing this reference to my attention.} and it can be related to solutions of the Knizhnik-Zamolodchikov equations. Consider the KZ equation,
\be
\partial_z L(z) = \Bigl( \frac{x_0}{z} + \frac{x_1}{1-z} \Bigr) L(z)
\ee
for noncommuting variables $x_0$ and $x_1$. The solution can be written
\be
L_X(z) = \sum_{w \in X^*} H(w;z) w
\ee
where $X^*$ is the set of all non-commuting words built from the alphabet $X=\{x_0, x_1\}$. In the arguments of the harmonic polylogs we treat the letters $x_0$ and $x_1$ as the letters $0$ and $1$ from section \ref{sect-shuffle}. Now consider
\be
L(z,\bar z) = L_X(z) \tilde{L}_X(\bar z)
\ee
where $\tilde{L}_X(\bar z)$ is the series with reversed words
\be
\tilde{L}_X(\bar z) = \sum_{w \in X^*} H(w;\bar z) \tilde{w}\,.
\ee
If we define
\be
L(z,\bar z) = \sum_{w \in X^*} L_w(z,\bar z) w
\ee
then the coefficients $L_w$ in this expansion (once antisymmetrised in $z$ and $\bar z$) correspond to the functions $f^{\rm top}_m(z,\bar z)$ for $w=m01S(m)$. The behaviour of the discontinuities around $z=0$ and $z=1$ can be seen \cite{shufflealg} from the fact that if we take $z$ around the point $0$ then $L(z)$ transforms to
\be
\mathcal{M}_0 L_X(z) = L_X(z) e^{2 \pi i x_0},
\ee
while if we take $z$ around the point $1$ we find
\be
\mathcal{M}_1 L_X(z) = L_X(z) Z_X^{-1} e^{-2 \pi i x_1} Z_X
\ee
where $Z_X = L_X(1)$, regularised with the shuffle relations. We can see, therefore, that around $z=0$, $L(z,\bar z)$ transforms to itself, i.e. is single-valued, in agreement with the analogous statement about $f^{\rm top}(z,\bar z)$ in section \ref{sect-solving}. Around $z=1$ we see that $L(z,\bar z)$ transforms to 
\be
\mathcal{M}_1 L(z,\bar z) = L_X(z) Z_X^{-1} e^{-2\pi i x_1} Z_X \tilde{Z}_X e^{2 \pi i x_1} \tilde{Z}_X^{-1} \tilde{L}_X(\bar z)\,,
\ee
and hence is not single-valued around $z=1$.

In \cite{fbsvpl}, single-valued polylogarithms were constructed by replacing $\tilde{L}_X(\bar z)$ by $\tilde{L}_Y(\bar z)$ which is the same series as $\tilde{L}_X(\bar z)$ but built on another alphabet $Y=\{y_0,y_1\}$. Then imposing that the functions are single-valued around $z=0$ and $z=1$ implies the relations between the two alphabets,
\begin{align}
y_0 &= x_0\,, \\
\tilde{Z}_Y y_1 \tilde{Z}_Y^{-1}  &=  Z_X^{-1} x_1 Z_X\,.
\label{y1eq}
\end{align}
To expand on the above a little, let us write down the generating series up to degree 3. We have 
\begin{align}
Z_X = 1 &+ \zeta(2)(x_0 x_1 - x_1x_0) \notag \\
&+ \zeta(3) (x_0x_0x_1 -2 x_0 x_1 x_0 +x_1x_0x_0 +x_0x_1x_1 - 2x_1x_0x_1 + x_1x_1x_0) \notag \\
&+ \ldots
\label{Zseries}
\end{align}
The relation (\ref{y1eq}) can be solved perturbatively for $y_1$,
\be
y_1 = x_1 -2\zeta(3)(w_3x_1 - x_1 w_3) + \ldots
\ee
where $w_3$ is shorthand for the coefficient of $\zeta(3)$ in equation (\ref{Zseries}). The $\zeta(2)$ part has dropped out in the above equation due to the antisymmetry of the coefficient of $\zeta(2)$ in (\ref{Zseries}). It follows from the above that 
\be
\mathcal{L}(z ,\bar z) = L_X(z) \tilde{L}_Y(\bar z)
\ee 
is single-valued around both $z=0$ and $z=1$. Thus an alternative to the approach we have followed here is to start from the coeffcients $\mathcal{L}_w(z,\bar z)$ defined by
\be
\mathcal{L}(z,\bar z) = \sum_{w \in X^*} \mathcal{L}_w(z,\bar z) w\,,
\ee
and build solutions of the differential equation (\ref{dezzb}) from them. Note that unlike for $L(z,\bar z)$ the differential equation we want is not automatic because the action of $\partial_{\bar z}$ introduces the $y$ letters on the right instead of the $x$ letters. Interestingly, this space of functions is also relevant for the discussion of the Regge limit of the six-particle scattering amplitudes of $\mathcal{N}=4$ super Yang-Mills theory \cite{Dixon:2012yy}.

\section{A conformal generalisation}
\label{sect-4ptconf}

\begin{figure}
\centerline{{\epsfysize8cm \epsfbox{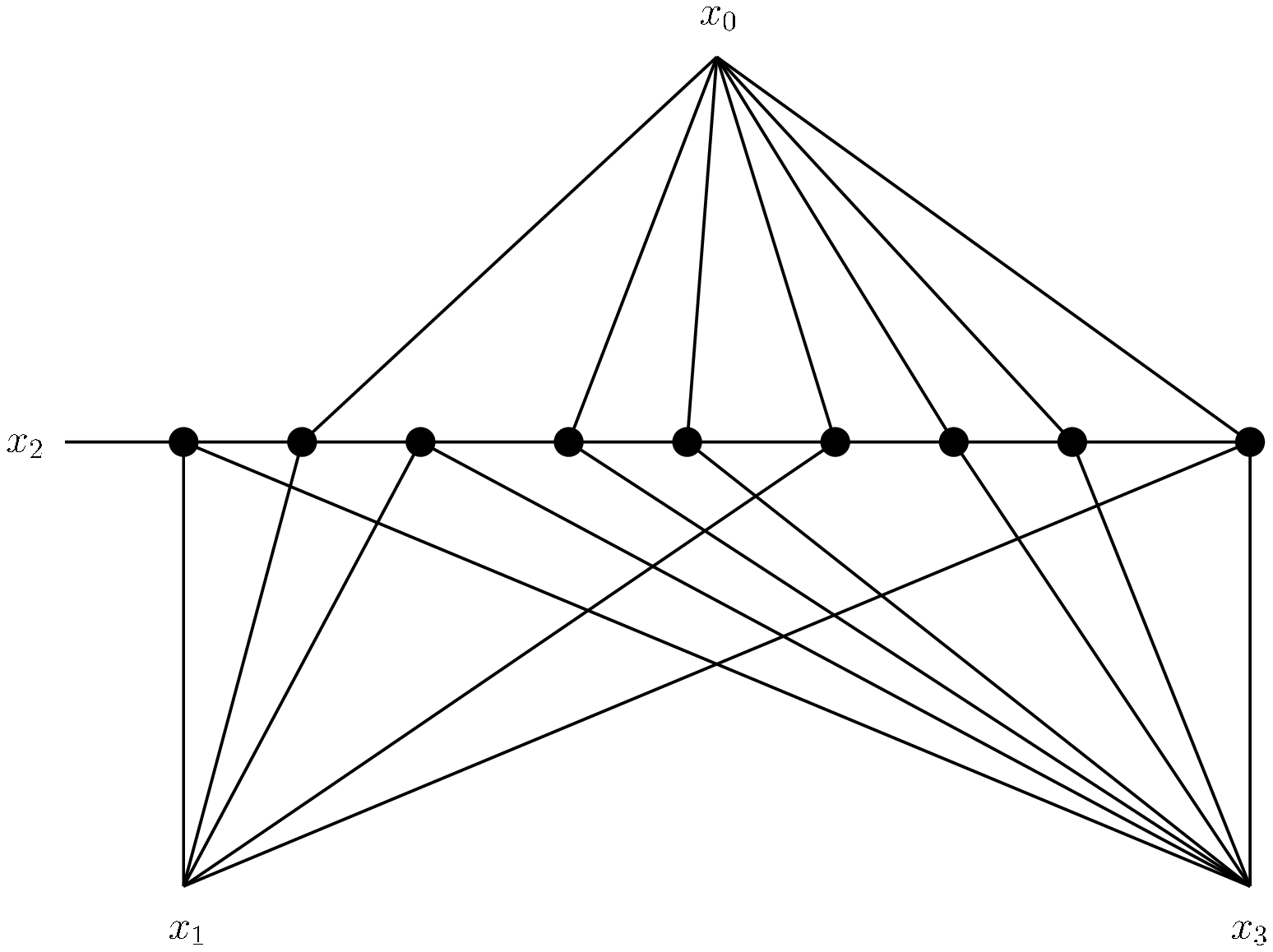}}} \caption[]{\small Conformal four-point integrals which generalise the three-point ladders introduced in section \ref{sect-genlad}. Taking the limit $x_3 \rightarrow \infty$ yields generalised ladders with both three-point and four-point integration vertices.}
\label{conformal4pt}
\end{figure}

We began the discussion with four-point ladder integrals whose dual diagrams exhibit conformal symmetry. To simplify the diagrams we took the limit where one of the four points was sent to infinity. We may of course restore the fourth point for the generalised ladder integrals. Doing so we obtain a four point integral with the fourth vertex $x_3$ attached to all of the integration points which are in turn attached to either $x_0$ or $x_1$. This suggests a natural generalisation where each integration point is attached to any two out of the three point $x_0$, $x_1$ and $x_3$. Such four-point integrals are depicted in Fig. \ref{conformal4pt}. We keep the convention that the integrand associated to the graph in Fig.\ref{conformal4pt} has numerator factors $x_{ij}^2$ depending only on external points such that the conformal weight at all four external points is 1. There is a unique assignment of such numerators which ensures this property.

Sending the point $x_3$ to infinity yields a class of three-point integrals where each integration point is attached to $x_0$ or $x_1$ or both. Such integrals satisfy differential equations of exactly the same form as described in section \ref{sect-genlad}. The only difference is that, in the case where the leftmost vertex is attached to both $x_0$ and $x_1$, we obtain an equation of the form
\be
\Box_2 I_m(x_0,x_1,x_2) = -\frac{4 x_{01}^2 }{x_{0 2}^2 x_{12}^2} I_{m'}(x_0,x_1,x_2)\,.
\ee
In terms of the associated functions $f_m$, we find the equations read
\be
z \bar{z} (1-z)(1-\bar{z}) \partial_z \partial_{\bar{z}} f_m(z,\bar{z}) = - f_{m'}(z,\bar z)\,.
\label{4ptintde}
\ee
Taking the prefactor to the right and using partial fractions we see that we can write the solution as a sum of four terms. It thus makes sense to associate an integration vertex which is attached to both $x_0$ and $x_1$ to the linear combination of letters $0+1$. The associated word $m$ is then given by $m= (0+1)m' = 0m' + 1m'$. The general approach outlined in the previous sections then goes through with only minor adaptations. In particular the top part is given by the same formula (\ref{ftopdef}) with the word $w$ built in the same way from the word $m$, i.e. $w = m 01 S(m)$ (hence the four terms in the differential equation (\ref{4ptintde}) after using partial fractions).

\section{Two-point limits and vacuum graphs}
\label{sect-vacuum}

We can obtain a two-point integral from one of our three-point integrals by taking a limit where $x_2$ approaches either $x_0$ or $x_1$. Such a limit will be finite as long as it does not produce a doubled propagator. Thus in the case of the generalised ladders described in Sect. \ref{sect-genlad} the limit $x_2 \longrightarrow x_1$ is finite in the case $a_1 = 0$ as is the limit $x_2 \longrightarrow x_0$ in the case $a_1=1$.

Let us consider the case $a_1=0$ so that $m=0m'$. We have
\be
\frac{V_{m'}}{x_{01}^2}= I_{m}(x_0,x_1,x_1) = \frac{1}{x_{01}^2} \lim_{x_2 \rightarrow x_1} v F_m(u,v) = \frac{1}{x_{01}^2} \lim_{z,\bar{z} \rightarrow1} \frac{f_{m}(z,\bar z)}{z-\bar z}= \frac{1}{x_{01}^2} \partial_z f_{m}(1,1)\,.
\ee
The final equality gives a simple formula for the normalisation of the two-point function,
\be
V_{m'} = \partial_z f_m(1,1)\,.
\label{Vdat1}
\ee
Similarly in the case $a_1=1$ we have $m=1m'$ and
\be
V_{m'} = \partial_z f_{m}(0,0)\,.
\ee
It is clear from the graph that the normalisation of the two-point integral does not depend on the first letter of $m$, hence the notation $V_{m'}$ above. Moreover there is also a reflection about the vertical axis of the graph leading to $V_{m} = V_{\tilde{m}}$ and a reflection on the horizontal axis leading to
$V_m = V_{\check{m}}$, where $\check{m}$ is the word $m$ with zeros and ones interchanged.

The same number is the value of the residue of the vacuum graph obtained from taking the planar dual of the two-point graph and joining the incoming and outgoing momentum lines. Note that the vacuum graph gains an additional loop when one glues the incoming and outgoing lines and so has $L+1$ loops.

\begin{figure}
\centerline{{\epsfysize10cm \epsfbox{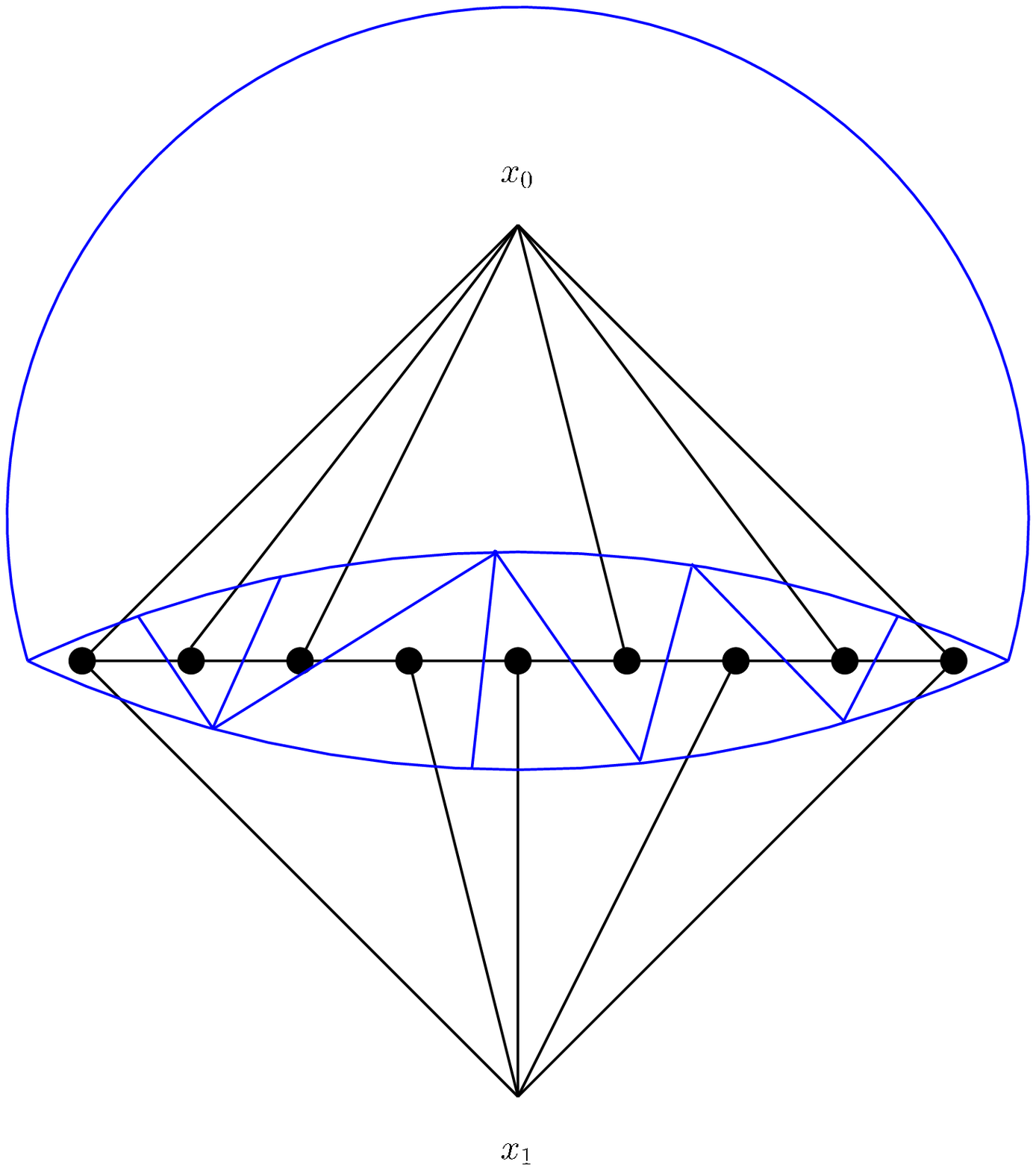}}} \caption[]{\small The finite two-point limit and the associated dual vacuum graph.}
\label{Fig:aux}
\end{figure}

The vacuum residues associated to the ladders give the wheel series,
\be
V_{0_{L-2}} = W_{L+1}\,.
\ee
where
\be
W_n = 
{2n-2 \choose n-1}
\zeta(2n-3)\,.
\label{Wn}
\ee

The wheel diagrams are also reproduced by the depth 3 generalised ladders, considered in section \ref{sect-depth3}, in the case $L=p+2$. In fact one can easily see that the only contribution to (\ref{firstd3int}) that survives in the limit $z\rightarrow0$, $\bar{z} \rightarrow 0$ is the the coefficient of $H_1(z)$ in (\ref{firstd3h}) which indeed reproduces the wheel formula above.

The full depth 3 series for $L>p+2$  gives the following vacuum graphs,
\begin{align}
V_{0_{L-p-3},10_p} = (-1)^{L-1} \biggl[ &\zeta_{L-p-2,p+2,p+1,0_{L-p-2}} - \zeta_{L-p-1,p+2,p+1,0_{L-p-3}} \notag \\
&+\sum_{r=0}^{p} \Bigl( \zeta_{L-p-2,p+2,0_{p-r}} \zeta_{L-p-1,0_r} - \zeta_{L-p-1,p+2,0_r} \zeta_{L-p-2,0_r}\Bigr) \notag \\
&+\sum_{r=0}^{p+1}\Bigl(\zeta_{L-p-2,0_r} \zeta_{L-p-1,p+1,0_{p+1,r}} - \zeta_{L-p-1,0_r}\zeta_{L-p-2,p+1,0_{p+1-r}}\Bigr) \notag \\
&+\zeta_{L-p-1,p+1,p+2,0_{L-p-3}} - \zeta_{L-p-2,p+1,p+2,0_{L-p-2}}\Bigr) \notag \\
&+\sum_{k=0}^{L-p-2} \Bigl( \zeta_{L-p-2,p+2,0_{L-p-2-k}} c_{p,k} + \zeta_{L-p-2,p+1,0_{L-p-2-k}}d_{p,k} \notag \\
& \qquad +\sum_{m=1}^{\lfloor \tfrac{p}{2} \rfloor} \zeta_{L-p-2,2m,0_{L-p-2-k}} e_{p,k,m} + \zeta_{L-p-2,0_{L-p-2-k}} t_{p,k} \Bigr)
\biggr]\,.
\end{align}
The top four lines come from the derivative of $f^{\rm top}_{L-p-1,0_p}$ while the final two lines come from the derivative of $f^{\rm ex}_{L-p-1,0_p}$. The explicit form of the coefficients $c,d,e,t$ was given in (\ref{cdetcoeffs}). Note that the only multi-zeta values which are not products of lower weight zetas come from the contribution of $f^{\rm top}$. The multi-zeta values with trailing zeros can be expressed in terms of those without by using the shuffle relations.
The depth 3 vacuum graphs are related to the G-graphs of  \cite{Broadhurst:1995km}.

The series $I_m$ where $m$ is an alternating sequence of ones and zeros is of interest because its coincidence limit reproduces the zigzag series of vacuum graphs (after taking the dual and closing the loop). For even $L$ we have
\be
V_{0,1,0,\ldots,0,1} = V_{1,0,1,\ldots,1,0} = Z_{L+1},
\ee
while for odd $L$ we have
\be
V_{1,0,1,\dots,0,1} = V_{0,1,0,\ldots,1,0}= Z_{L+1}\,.
\ee
The zigzag vacuum graphs were conjectured in \cite{Broadhurst:1995km} to be given by
\be
Z_n = \frac{4 (2n-2)!}{n! (n-1)!}\biggl(1-\frac{(-1)^n}{2^{2n-3}}\biggr) \zeta(2n-3)\,.
\label{zigzags}
\ee
Using the methods described in section \ref{sect-svb} we have verified this conjecture to ten digit precision\footnote{Although higher precisions are often required to distinguish different linear combinations of multi-zetas, here we are testing a conjectured simple zeta with a precise rational coefficient so this precision seems perfectly adequate for our purposes.} up to 13 loops for the vacuum graphs, i.e transcendental weight 23. Note that, despite the fact that the zigzag graphs are of very high depth, the resulting vacuum residues are conjecturally simple zetas only. In fact we observe a similar simplicity in constructing the single-valued polylogarithms for the generalised ladders where $m$ is an alternating sequence of zeros and ones; the only explicit multi-zeta values which appear are odd simple zetas for odd weights and products of two odd simple zetas for even weights.

The generalised ladders give a large family of vacuum graphs that contains both the wheel and zigzag series. Such vacuum graphs have been called generalised zigzags \cite{Dorynthesis}. The discussion above and of section \ref{sect-gensvpolys} shows that we have related all these graphs to the derivatives at $z=1$ (or $z=0$) of the single-valued polylogarithms associated to the words $w$ describing the generalised ladder graphs. From section \ref{sect-solving} we have seen that these functions can be described in terms of a top part, which can be read off straightforwardly from the graph, and an extra part which involves explicit multi-zeta values.  From (\ref{Vdat1}) we see that
\be
V_{m'} = \partial_z f_{0m'}(1,1) = \partial_z f^{\rm top}_{0m'}(1,1) + \partial_z f^{\rm ex}_{0m'}(1,1)\,.
\ee

Almost all terms in the above formula are products of multi-zetas of lower weights. The only terms which are not immediately products come from the contribution due to $f^{\rm top}$ and are given the extreme and next-to-extreme terms in the deconcatenation coproduct of the word $w$ in equation (\ref{ftopdef}). They are given explicitly by the formula,
\be
V_{m'} = (-1)^{L-1}\bigl[ \zeta_{m'01\tilde{m'}0} - \zeta_{m' 1 0 \tilde{m'} 0} + \zeta_{0 m' 10\tilde{m'}} - \zeta_{0m'01\tilde{m'}}\bigr] + \text{ products.}
\ee
Using the fact that 
\be
\zeta(w) = -\zeta(S(w)) + \text{ products,}
\ee
we find
\be
V_{m'} = 2(-1)^L\bigl[\zeta_{0m'01\tilde{m'}} - \zeta_{0m'10\tilde{m'}}\bigr] + \text{ products.}
\ee
In the case of the zigzag graphs with $L$ even we have 
\be
V_{1,0,1,0,\dots,0} = 2\bigl[ \zeta_{2_{L/2 -1} 3 2_{L/2-1}}  - \zeta_{2_{L/2} 3 2_{L/2-2}} \bigr] + \text{ products,}
\ee
where $2_r$ denotes a string of $r$ twos. Schnetz has observed \cite{os} that if we drop all product terms, the result of Zagier \cite{zagier} implies the result (\ref{zigzags}). Zagier's formula tells us\footnote{Note that our conventions for the ordering of the labels in the multi-zeta values differ from those of \cite{zagier}.}
\begin{align}
\zeta_{2_b 3 2_a} = &2 (-1)^{a+b+1}\biggl[\bin{2(a+b+1)}{2a+2} - \Bigl(1 - \frac{1}{2^{2(a+b+1)}}\Bigr)\bin{2(a+b+1)}{2b+1}\biggr]\zeta_{2(a+b+1)+1} \notag \\
&+ \text{ products}\,,
\end{align}
which, leads to (recall $L$ is even)
\be
V_{1,0,1,0,\dots,0} =  Z_{L+1} + \text{ products.}
\ee
In agreement with (\ref{zigzags}). Similarly if $L$ is odd we obtain
\be
V_{1,0,1,0,\dots,1} = 2\bigl[ \zeta_{2_{(L-1)/2 } 3 2_{(L-3)/2}}  - \zeta_{2_{(L-3)/2} 3 2_{(L-1)/2}} \bigr] + \text{ products}\,,
\ee
which implies
\be
V_{1,0,1,0,\ldots,1} = Z_{L+1} + \text{ products.}
\ee
Thus the test of (\ref{zigzags}) described above is essentially a test that the product terms actually vanish. It remains to show analytically that the products do indeed vanish to all loops. It is reasonable to hope that the methods described in this paper will allow for such a proof.

\begin{figure}
\centerline{{\epsfysize8cm \epsfbox{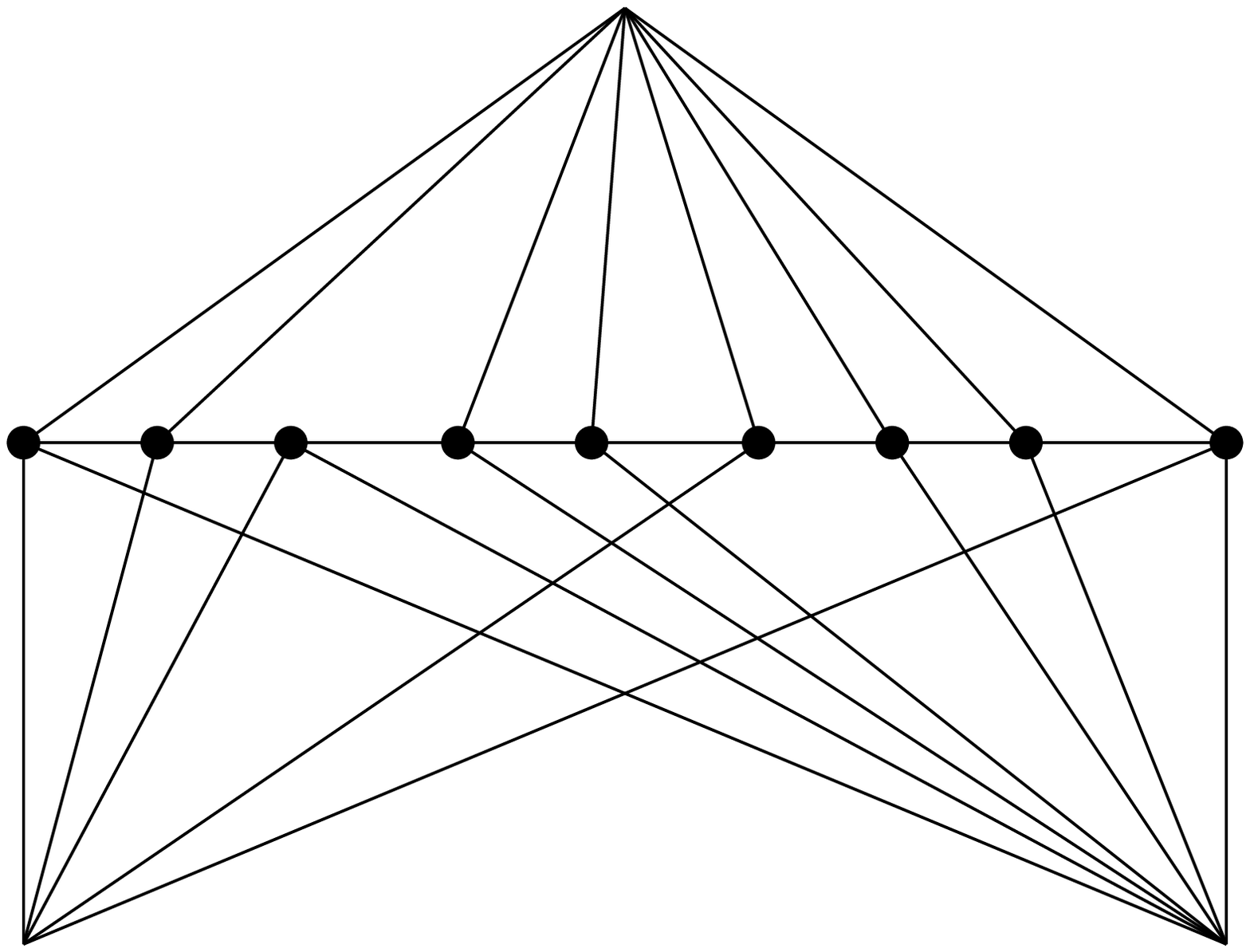}}} \caption[]{\small The conformal three-point stars that can be obtained by a finite limit from the four-point conformal integrals of section \ref{sect-4ptconf}.}
\label{conf3pt}
\end{figure}

Given the considerations of section \ref{sect-4ptconf} we can state more generally that we can apply the methods described here to derive the value of all three-point conformally covariant graphs which have all integration vertices arranged in a line and no numerators (see Fig. \ref{conf3pt}).

\section{Summary}

We have discussed the class of generalised ladder integrals and the differential equations that they satisfy. The ambiguity in solving the differential equation is fixed by the fact that the solutions are given by single-valued polylogarithms, which generalise the Bloch-Wigner dilogarithm. The symbols of the corresponding functions can be immediately related to the associated integrands via some simple operations which arise naturally from the shuffle Hopf algebra. The `top part' or, equivalently, the symbol, is related in a simple way to a solution of the Knizhnik-Zamolodchikov equation.
The additional terms involving explicit multi-zeta values can be fixed recursively by imposing single-valued behaviour on the solutions of the differential equations. We have given an explicit construction of an $L$-loop function from an $(L-1)$-loop function. In particular we have given an all-loop formula for the depth 3 integrals, the simplest integrals in the class which are not ladders.

The results for the generalised ladders can be applied to derive an infinite class of vacuum diagrams, the generalised zigzags, or more generally, following the relation to conformal four-point integrals, all conformal three-point stars with a linear integration vertex topology and no numerators. In particular we have tested the conjecture of Broadhurst and Kreimer for the form of the zigzag diagrams and offer a method for deriving that series of diagrams.

\subsubsection*{Acknowledgements}

I would like to thank Francis Brown for discussions and Oliver Schnetz for confirming some low-loop results for the depth 3 vacuum graphs in the early stages of this work.

\appendix

\section{Discontinuities of harmonic polylogarithms}
\label{app-Fmap}

Here we describe the map $\mathcal{F}$ we used in constructing the holomorphic function $h(z)$ in section \ref{sect-svb}. The map takes a linear combination of harmonic polylogs to another combination of harmonic polylogs with no discontinuity around $z=0$ but without changing the discontinuity around $z=1$,
\be
{\rm disc}_z \mathcal{F}[f(z)] =0\,, \qquad {\rm disc}_{1-z} \mathcal{F}[f(z)] = {\rm disc}_{1-z} f(z)\,.
\ee
On harmonic polylogs whose defining word ends in a 1, the map is the identity
\be
\mathcal{F}[H(w1;z)]=H(w1;z)\,.
\ee
For harmonic polylogs whose defining word ends with a string of zeros we have
\be
\mathcal{F}[H(r_1,\ldots,r_d,0_p;z)] = \sum_{k=1}^{d-1} H(r_1,\ldots,r_{d-k};z)F(r_{d-k+1},\ldots,r_d,0_p)\,.
\label{Fop}
\ee
Here $F$ is a combination of multi-zeta values defined recursively by
\be
F(s_1,\ldots,s_n,0_p) = \zeta(s_1,\ldots,s_n,0_p) - \sum_{k=1}^{n-1} \zeta(s_1,\ldots,s_k) F(s_{k+1},\ldots,s_n,0_p)\,,
\ee
while
\be
F(s,0_p) = H(s,0_p;1)=\zeta(s,0_p)\,.
\ee
The fact that ${\rm disc}_z \mathcal{F}[f(z)]$ vanishes is obvious since no harmonic polylogs on the RHS of (\ref{Fop}) have trailing zeros. The fact that the discontinuity around $z=1$ of $\mathcal{F}[f(z)]$ matches that of $f(z)$ can be justified recursively using (\ref{discomzg0},\ref{divdisc}).

\setcounter {equation} {0}

\end{document}